\documentclass[12pt,preprint]{aastex}

\def\referee#1{#1}

\def\breferee#1{#1}

\shorttitle{Eruptions from solar ephemeral regions}
\shortauthors{Schrijver}

\begin{document}

\title{Eruptions from solar ephemeral regions as an extension of the size distribution of coronal mass ejections}

\author{Carolus J. Schrijver}

\affil{Lockheed Martin Advanced Technology Center, Palo Alto, CA 94304}

\email{schryver@lmsal.com}

\begin{abstract}
Observations of the quiet solar corona in the 171\,\AA\ ($\sim 1$\,MK)
passband of the
Transition Region and Coronal Explorer (TRACE) often show
disruptions of the coronal part of small-scale ephemeral bipolar regions that
resemble the phenomena associated with coronal mass ejections on 
much larger scales: ephemeral regions exhibit
flare-like brightenings, rapidly rising filaments 
carrying absorbing material at chromospheric temperatures, or the
temporary dimming of the surrounding corona. I analyze all available
TRACE observing sequences between 1998/04/01 and 2009/09/30 with
full-resolution 171\,\AA\ image sequences spanning a day or more within
500\,arcsec of disk center, observing essentially quiet Sun with 
good exposures and relatively low background. Ten such data sets
are identified between 2000 and 2008, spanning 570\,h of observing
with a total of 17133 exposures. Eighty small-scale
coronal eruptions 
are identified. Their size distribution forms a smooth extension
of the distribution of angular widths of coronal mass ejections, 
suggesting that the eruption frequency for
bipolar magnetic regions is essentially scale free over at least
two orders of magnitude, from eruptions near the arcsecond resolution limit of
TRACE to the largest
coronal mass ejections observed in the inner heliosphere.  
\referee{This scale range may be associated with the
properties of the nested set
of ranges of connectivity in the magnetic field, in which increasingly
large and energetic events can reach higher and higher into the corona
until the heliosphere is reached.}
\end{abstract}

\keywords{Sun: corona --- Sun: coronal mass ejections (CMEs) --- Sun: magnetic fields}

\section{Introduction} 

Eruptive and explosive events in the solar corona exhibit a tendency
for self-similar behavior that expresses itself in power-law
distributions of frequency versus, e.g., size, total energy, or peak
brightness \citep[e.g.,][and references
therein]{1971SoPh...16..152D,crosby+etal1993,aschwanden+etal2000b}. In
the case of solar flares, there is a remarkable scaling from small
flares observed in the EUV to large flares seen in hard X-rays, with
essentially the same power-law index describing quiet-Sun
'nanoflares', active-region transient brightenings, and hard X-ray
flares over eight orders of magnitude in estimated flare energy
\citep{aschwanden+parnell2002}.  A similar power-law behavior,
albeit observable over a much smaller range in total energies, has been
reported on for the energy distribution of large stellar flares
\citep{audard+etal2000a}. A power law is also a good approximation to,
for example,
the distribution of area or flux in recently emerged active regions
\citep{harvey+zwaan93}, extending relatively smoothly into the domain of
ephemeral regions \citep{hagenaar+etal2003} over almost five orders of
magnitude in absolute magnetic flux.

In a study of SOHO/LASCO coronagraphic 
observations of the inner heliosphere, \citet{robbrecht+etal2009} present
evidence that the angular width of coronal mass ejections (CMEs)
exhibits a scale-free power-law distribution that extends from
about 20$^\circ$ up to 120$^\circ$, i.e. over a range of a factor
of about six in opening angle. This result is based on 
the use of an automated feature detection method, CACTus, applied
to LASCO observations from September 1997 through January 2007.
Earlier visual inspection of the LASCO data had suggested that 
the CME distribution peaks at an angular width of about 30$^\circ$, but
the CACTus software also identifies many smaller structures. On
visual inspection by \citet{robbrecht+etal2009} these are seen to 
fall into several categories, including visually-identified 
events that are broken up
by the CACTus algorithm, trailing outflows, wavelike phenomena, slowly rising
loop-like structures, opening field, and some 'false detections.' 
It is unclear whether most of these features should be classified
as true CMEs, but the present study puts these results in 
a perspective that suggests that 
perhaps these relatively narrow events identified by CACTus 
(or at least many of them) may well be part of a scale-free continuous
distribution of eruptive events.

The apparently scale-free CME frequency 
distribution function over a factor of about 6 in
angular width as found by the CACTus software begs the question as
to what happens on even smaller scales. Here, not only the spatio-temporal
resolution of the available telescopes comes into play, but also the
very real possibility that small-scale, and generally less-energetic, eruptive
events may not be able to escape through the overlying coronal field,
and thus never develop into a proper CME while possibly having comparable
properties for their associated field eruptions lower down
(but see, e.g.,  Mandrini et al.\ (2005)\nocite{mandrini+etal2005} 
for an example of a very small
eruption that they argue did make it into the 
heliosphere, \breferee{and Wang and Sheeley (2002)\nocite{wang+sheeley2002} for
a discussion of narrow SOHO/LASCO jets in or near coronal holes).}

In order
to investigate the statistics of small-scale coronal eruptions that
are characteristically, but not uniquely, associated with ephemeral
bipolar regions in the solar photosphere, I investige a sample
of ten data sets obtained by the TRACE telescope between 2000 and
2008, and compare the results to the CME statistics as derived
by \citet{robbrecht+etal2009}.

\referee{Ephemeral regions are small bipolar magnetic regions that
contain no sunspots or pores. Their unsigned magnetic fluxes range up
to about $10^{20}$\,Mx, above which bipolar regions are generally
called active regions. At the small end of their flux spectrum, they
extend into the intranetwork mixed-polarity field; a rather vaguely
defined lower limit to their unsigned magnetic flux of order $3\times
10^{18}$\,Mx is sometimes used, but no flux range has been formally
defined for ephemeral regions. These regions behave like small active
regions on initial emergence, with the two polarities fragmented into
a set of smaller flux clusters that quickly separate on emergence,
often showing complex meandering motions suggesting a tangled emerging
field.  Shortly after emergence, they are subjected to the
supergranular flow, and the polarities drift into the network lanes,
subject to canceling collisions or mergings with other network
elements that are already there.  They emerge almost uniformly across
the solar surface, show little dependence on the solar cycle, and have
an essentially random orientation altough the larger ones seem to have
a slight preference for the dipole-axis orientation proper for the
dominant magnetic cycle, and their emergence frequency decreases with
increasing flux imbalance within the surrounding photosphere as seen
in unipolar regions formed by decaying active regions. The
distribution of their emergence frequency as a function of unsigned
flux appears to be a smooth extension of the near-power-law
distribution found for active regions, suggesting that ephemeral
regions are a transition population between the large cycle-related
active regions and the apparently ubiquitous mixed-polarity background
of the internetwork field \citep[see, e.g.,][and references
therein]{harvey+zwaan93,schrijver+zwaan99,hagenaar+etal2003,hagenaar+etal2008}.}

\section{Observations}\label{sec:features}
The observations of the small-scale coronal eruptions were
obtained with the Transition Region and Coronal Explorer
\citep[TRACE;][]{traceinstrument} in its 171\,\AA\ passband, which
has a peak sensitivity around 1\,MK. From the mission archive
from 1998/04/01 through 2009/09/30, I selected all data sets
pointing at quiet Sun, within 500\,arcsec of disk center, extending
over at least one day, with 171\,\AA\ images as the primary
observing passband, and with exposures mostly exceeding 42\,s in duration
in order to ensure an adequate signal to noise ratio. The initial
selection of potentially suitable data sets was facilitated by a
visual mission summary made publically available via
the TRACE mission home page.\footnote{URL: http://trace.lmsal.com/POD/TRACEpodarchive28.html{\#}case02}

These candidate data sets were then inspected for data quality,
selecting only those sets with generally good exposures with
acceptable background levels and radiation damage, that as a set have
few or only short interruptions, and with a field of view between
$4.25\times 4.25$ and the full $8.5\times 8.5$ arcminutes squared
(thus, for example, generally excluding observations in the
November--January orbital 'eclipse season' because of the frequent
transits of the Earth and the relatively high instrumental background
level owing to a raised detector temperature in that phase of the
year). The remaining ten data sets, with a total of 17133 exposures,
are listed in Table~\ref{tab:statistics}. The effective duty cycle
(estimated by identifying intervals with useful observations with
interruptions of 30\,min.\ or less) for each of these data sets
exceeds 89\%.

The image sequences were analyzed visually by displaying the series
with normalized intensity scalings, corrected for particle hits on the
detector ('despiked'), and by tracking the region for solar rotation,
while offsetting for instrumental pointing changes. Events were
selected that resembled small equivalents of active-region eruptions
associated with CMEs, specifically looking for (a) erupting dark
fibrils (like erupting filaments), (b) rapid dimmings around a compact
ephemeral region (equivalents of large-scale coronal dimmings), or (c)
very rapid reconfigurations of a mix of dark and bright coronal
structures above ephemeral regions often linking to one or more
neighboring regions.

\referee{For each of the selected events with one or more of the
mentioned characteristics, I measured the largest length scale over 
which the perturbation of the corona was apparent, thus measuring the
extent over which the eruption unfolds rather than the extent of the original
source region or any particular one of the three abovementioned
characteristics. I do not differentiate
between the three characteristics as many events display them in
conjuction. I return to this point in \S~\ref{sec:discussion}.}

Many other events occur within the complex, dynamic quiet-Sun corona
\referee{in addition to those selected by the above criteria. Their inclusion
in, or exclusion from, the currently discussed sample is admittedly
subjective. Excluded were, for example, compact, contained flare-like
brightenings, many of which are associated with very narrow, jet-like
cusps.  On the other hand, events with clearly eruptive signatures
that included a jet-like event, for example, were included.  This selection is
similar to the distinction between larger GOES-class flares and CMEs:
many (particularly the largest) flares are associated with CMEs, but
in this study I focus on events that have clear signatures of a
disruption of the magnetic field of the ephemeral region.}

Figure~1 shows select examples of the types of events studied in
here. Rows {\em a} and {\em b} show very small, 
bubble-like eruptions (marked by arrowheads in one of the frames for 
each case), 
in which a one-sided dimming suddenly appears; the size of the 
coronal dimming for case {\em a} in the image taken at 21:56\,UT
corresponds to 10 \referee{times the 1-arcsec TRACE resolution}, which puts it at the limit of what
can be reliably interpreted as an eruption based on the TRACE data.
Row {\em c} shows a much larger example of such an event, with a clear coronal
dimming extending over approximately 100\,arcsec. Rows {\em d}
and {\em e} show eruptions in which a two-sided dimming is seen
(marked by the arrow heads),
as sometimes observed in association with an active-region eruption.
Row {\em f} shows an event with a dimming as well as a small,
dark fibril (indicated in the 15:50UT frame by the arrowhead). Row {\em g} 
shows an example of a very compact eruption in association with a flare
that is so bright that the diffraction pattern caused by the filter
support grid shows up, as seen in the 2nd and 3rd panels. Row {\em h} shows
an eruption in which the ephemeral region field connects to two relatively
distant neighboring flux concentrations, extending in fact beyond the
shown cutout of the full field of view of the observations. 

The examples shown in Figure~1 are characteristic in their appearance,
although they were selected from  
image sets that are well exposed with little damage
by background radiation by energetic particles or by a high readout
noise associated with relatively high detector temperatures in some
phases of the orbits (depending on the season in which the observations
are obtained). The diversity of phenomena, the frequent short-term 
interruptions of good observing conditions by energetic-particle
impacts or Earth-atmospheric absorption, and the fact that we are
evaluating pattern evolution rather than curves (such as flare
brightness curves) severely hamper the ready 
application of an automated feature
finding algorithm, instead requiring a visual identification of the events with
the associated subjectivity of such a procedure. 

The set of all images, covering 570\,h of observing, yielded 80
events that resembled small-scale active-region eruptions as defined
above (see Table~\ref{tab:statistics}). Most of these events originate
in the bright corona over ephemeral regions, but some occur in the
connections between such regions, while two occur within the
largely unstructured regions over very quiet Sun. The estimated event duration
ranges from 3\,min.\ to 104\,min., with an average and
standard deviation of $33\pm 20$\,min. For each eruption a 
characteristic length scale
was estimated using the maximum extent of the event in any direction
as measured on the images; these
length scales range from 6\,Mm to 160\,Mm, with an average and
standard deviation of $28\pm 24$\,Mm. The average histogram of
the event size distribution, normalized to events per day on 
the Sun under the assumption of a homogeneous surface distribution, 
is shown in Fig.~2, binned into intervals with a width of a factor of two.
\referee{The assumption of a uniform distribution over the solar surface
holds only to first approximation: \citet{hagenaar+etal2003}
show that the latitude distribution at least up to about 60$^\circ$
-~the limit of the range included in their analysis~- is 
somewhat tapered towards higher latitudes, but the distribution is
broad and featureless and the assumption of uniformity good to
within about a factor of two, commensurate with the statistical
uncertainties of the number of events analyzed here.}

\section{Discussion and conclusions}\label{sec:discussion}
The distribution of size scales for eruptive disturbances in the
quiet-Sun corona (Fig.~2) shows a pronounced decrease
in frequency with increasing size above the interval for the smallest 
selected events.
The apparent turnover at the smallest scales is likely an artefact of the
instrumental resolution: events smaller than about 10 resolution elements
are difficult to assess, and identifying an erupting small filament or
a coronal dimming for such compact features is 
problematic, so that these small events are likely significantly
underrepresented in the sample. The interval for the largest eruptions 
contains only two events, both from the same region, so that the uncertainty
on its frequency is substantial. 

The statistics for the
individual time intervals for the three remaining size intervals do
not suffice by themselves to say much about the shape of the distribution
function as a function of the phase of the solar cycle.
Even with all data combined, the slope of the size distribution is
rather poorly constrained based on the TRACE data by themselves. 
\referee{I propose, however, that the empirical evidence
suggests the combination of the events identified in the present study 
with the results on CME widths
from the study of LASCO observations
by \citet{robbrecht+etal2009}. I argue that in both cases, i.e.,
for the small eruptive events in quiet Sun and for global CMEs, 
the field somehow destabilizes, erupts, and is at least temporarily
disrupted. In the case of the quiet-Sun
eruptions of ephemeral-region field studied here, the ultimate
extent of the event is
likely restricted to the range of the set of magnetic
connections of a bipolar region, unless it is energetic enough to breach
that and reach into the next set of the overlying hierarchy. The same
argument can be made for eruptions in the coronal field over
active regions or over large quiet-Sun filaments, although here
the highest domain of connectivity obviously reaches into the heliosphere.
An illustrative example is shown in Fig.~3a, which mimics the potential
field in the so-called PFSS (potential-field source-surface) 
approximation in which the field is forced to become radial at
a distance of roughly 2.5 solar radii from the Sun's center 
\citep[see, e.g.,][for references and a comparison of the PFSS
model results with MHD simulations of the quiescent coronal-heliospheric
field]{riley+etal2006ApJ}. In this scenario, an eruption may be
associated with an expanding filament, and all or only part of 
the coronal region into which
the eruption unfolds may or may not exhibit a coronal dimming
\citep[see, for example,][for a comparison of the CME-related coronal dimmings
and the CME angular size]{reinard+biesecker2008}.
If, in the field configuration shown in Fig.~3a, a large eruption
would breach the high coronal field from the strong bipolar region
centered at an angle of 45$^\circ$ in the diagram, the associated
equivalent CME would have an opening angle into the heliosphere
close to that shown by the two dashed lines; if, however, the field
would contain the eruption and ejecta would be contained to within the
range of magnetic concentrations to which the central region is connected, 
a rather comparable angle would be spanned as projected onto 
the solar surface -~as sizes are expressed in this study
\citep[Fig.~1 in][also serves as an illustration in the literature
pertinent to 
this argument for a plane-parallel approximation valid for
scales small relative to the solar radius]{schrijver+title2003L}. 
Figure~3b illustrates how insensitive the inner coronal field is
to the upper boundary condition of radial field. This 
single, highly abstract example of a PFSS field extrapolation 
does not suffice for a general conclusion, of course, but 
the correspondence of quiet-Sun eruptions and CME statistics
may well point to this argument as the reason for their 
surprising alignment over a range of scales, as I now discuss.}

\referee{Figure~2 shows that the
average power-law fit to the events reported by \citet{robbrecht+etal2009}
for the overlapping period of
2000 to 2006 forms a continuous extension of 
the size distribution from the present study, albeit at a somewhat
steeper slope than the average fit to the CME distribution. Note
that Figure~2 shows the mean power-law fit from \citet{robbrecht+etal2009}
from an angular width of 20$^\circ$ upward, i.e., for the range in
their results in which a power-law shows a good fit to the observed
frequency distribution, and excluding -~as for the TRACE observations~-
very narrow jet-like events \breferee{(similarly, narrow
jet-like "collimated ejections" seen in LASCO observations 
\citep[e.g.][]{wang+sheeley2002} 
with width of up to $\sim 15^\circ$ are not included in the range
to which power-law fits were made to the CACTus CME data)}}.
Although the distribution of ephemeral region eruptions lies somewhat
above the average best fit to the CACTus CME distribution, they extend
that distribution within the range of fits seen in the period of 2000 to
2006. The CACTus CME distribution does not differentiate between events
originating in association with active regions or with quiet Sun, but 
\breferee{the CME distribution is dominated 
by events associated with active regions: \citet[][consistent with
earlier studies referenced therein]{zhou+etal2003},
for example, find that 79\%\ of front-side halo CMEs are associated
with activity within active regions.}

The results in Figure~2 suggest that the eruption frequency in large
bipolar regions (for CMEs associated with bipolar regions) 
and in their (quiet-Sun) surroundings is a scale-free quantity
that extends over a factor of almost 100 (but perhaps more), from the largest
coronal mass ejections to eruptions near the resolvable limit of
order 10 arcseconds for the highest-resolution EUV telescope. 

Our knowledge about what happens on even smaller length scales is, for
obvious reasons, rather limited.  In a recent study,
\citet{innes+etal2009} discuss what they call 'quiet Sun
mini-CMEs'. They analyze a sequence of 171\,\AA\ images obtained by
the STEREO-A spacecraft, with 150\,s cadence (comparable to the
average cadence in the data set discussed here) and 1.6\,arcsec pixels
(3.1 times larger than the TRACE pixels). They 
estimate a total number of events of 1400 per day for the entire
Sun when assuming a uniform surface distribution. 
They do not specifically 
count events equivalent to those reported on here  that
look like eruptions of small active-regions, but look for
"emission or absorption trains by eye in series of time-distance
171\,\AA\ images" and make "sensible choices for selecting events."
They do not provide a size distribution, or an average size, of their
events, but do note that "only events seen over 6$^{\prime\prime}$ (3
consecutive pixels) were kept." Assuming that the steep power law found here
continues to the smallest scales that they include, the number of smallest
events will dominate the total count. If we assume that these events
all reside in an interval with a width of a factor of two, as used for
the TRACE data in Fig.~2, extending from 6 to 12 arcseconds, then the
equivalent position of the results by \citet{innes+etal2009} is shown
by the gray bar in Fig.~2. As their count includes flare-like
brightenings, rapid coronal configurations, as well as the small
eruptions counted in the TRACE data analyzed here, the position of the
\citet{innes+etal2009} results is compatible with the TRACE
results, \referee{even} though the mix of event types in their study does not provide a
strong constraint on the extension of the power law to very small
scales. 

Altogether, the data from SOHO/LASCO, TRACE, and STEREO suggest
an essentially 
scale-free frequency distribution for sizes of magnetically driven eruptions
in the Sun that extends from the smallest scales that can  be
observed by present-day high-resolution instruments up to large-scale
coronal mass ejections. The grey dashed-dotted line in Figure~2 suggests
that this frequency distribution may be approximated to first order by 
a power-law distribution with an index of about $-2.3$ (to be
compared 
to the equivalent power-law index for length scales of loops involved 
in small-scale quiet-Sun
flaring of $-2.10 \pm 0.11$ reported by Aschwanden et al., 2000). 

\referee{The characteristic power-law index of $-2.3$
and the possible steepening when going from large to smaller scales 
has an intriguing analogy in the flux distribution of newly emerging
bipolar regions from large active regions to small ephemeral regions,
as summarized by Hagenaar et al.\ (2003) in their Fig.~11:
the apparently smooth transition from ephemeral to active regions
can be approximated by a single power law fit with a slope rather close
to $-2.3$. 
It will be interesting, in a future study, 
to explore in detail the reasons behind this
commonality, which includes at least assessing relationships between
the longevity of regions as a function of their size
\citep[e.g.,][]{harvey+zwaan93} and the evolution of 
their propensity to erupt during their
life time \citep[possibly related to the phenomenon of active-region
nesting, see][]{Brouwer+Zwaan1990,harvey+zwaan93}, and the
relationship between the properties of active regions and the
extent of their possible eruptions \citep[e.g.,][]{moore+etal2007}.}

Although the TRACE data sets studied here span a large part of the past 
sunspot cycle, the number of events detected per data set is too small
to make significant statements about the possible dependence of the number
of events on the phase of the sunspot cycle. Data sets at latitudes other
than near disk center are even rarer in the TRACE records. 
In time, the Atmospheric Imaging Assembly on the 
future Solar Dynamics Observatory should enable a more comprehensive study
of the statistics of erupting bipolar regions from ephemeral to active
regions both as a function of latitude and cycle phase. 

\acknowledgements 
I thank M.\ Aschwanden, B.\ De Pontieu, and N.\ Nitta for comments on the 
manuscript, and the referee for stimulating comments leading to an 
improved presentation of the methods and findings and an
expanded interpretation of the result. This work was supported by NASA
under the TRACE contract NAS5-38099 with NASA Goddard Space Flight
Center.


\vfill\eject
\begin{table}[ht]
\begin{center}
\caption{\em 
Summary of TRACE 171\,\AA\ data sets and the observed small-scale eruptive 
events.}\label{tab:statistics}
\begin{tabular}{lrrrrr}
\hline
Time interval & $\Delta t$ & Duty & fov  & no.\ & no.\  \\
& (h) & cycle(\%) & (armin$^2$) & images & events \\
\hline
2000-08-26 00:16UT to 2000-08-28 12:59UT& 60.8&92&42&1548&7\\
2003-03-25 00:08UT to 2003-20-26 23:59UT& 48.0&100&55&1236&10\\
2004-07-10 01:01UT to 2004-07-11 23:59UT& 47.0&100&16&560&2\\
2005-04-02 00:19UT to 2005-04-04 06:39UT& 54.0&95&54&1569&13\\
2006-08-04 09:57UT to 2006-08-07 23:59UT& 86.0&94&56&4690&22\\
2006-10-15 05:46UT to 2006-10-20 13:58UT&128.0&99&58&4701&8\\
2007-04-05 00:22UT to 2007-04-06 05:55UT& 29.6&100&36&609&4\\
2007-06-19 01:11UT to 2007-06-20 23:59UT& 46.9&89&58&672&5\\
2007-09-03 01:45UT to 2007-09-04 23:59UT& 46.2&100&36&870&5\\
2008-02-12 00:37UT to 2008-02-12 23:59UT& 23.4&100&58&678&6\\
\hline
\end{tabular}
\end{center}
\end{table}

\clearpage

\begin{figure}
\epsscale{.85}
\plotone{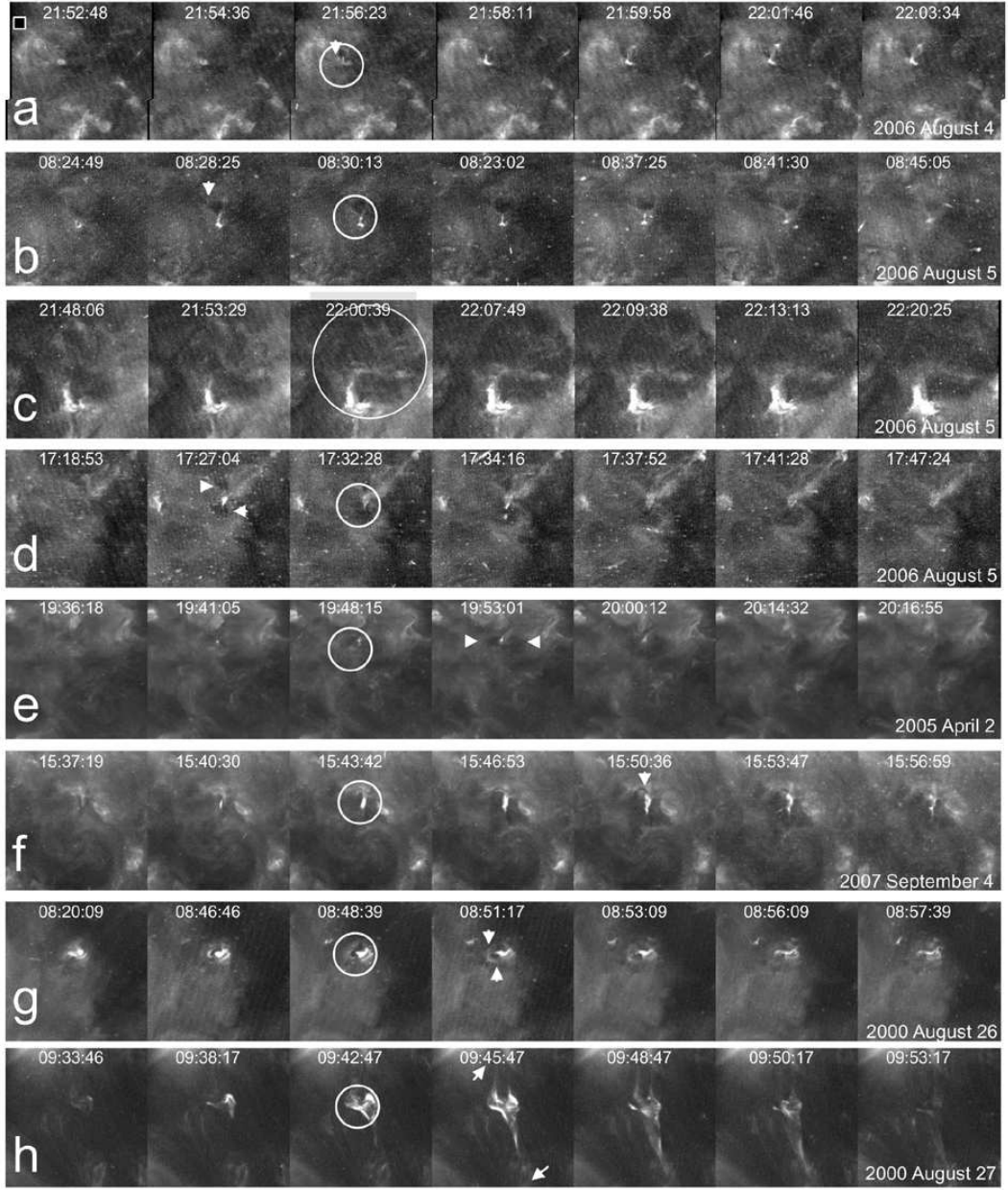}
\caption{Examples of events studied in this manuscript. Each event
is characterized by selecting seven TRACE exposures taken 
in the 171\,\AA\ pass band, each showing a field
of view of $120\times 120$\,arcsec (times are shown in UT).
Arrows and circles show characteristic features discussed in
\S~\ref{sec:features}. The small black square in the top-left
panel has a linear dimension of 10\,arcsec.}
\end{figure}
\clearpage

\begin{figure}
\epsscale{.47}
\plotone{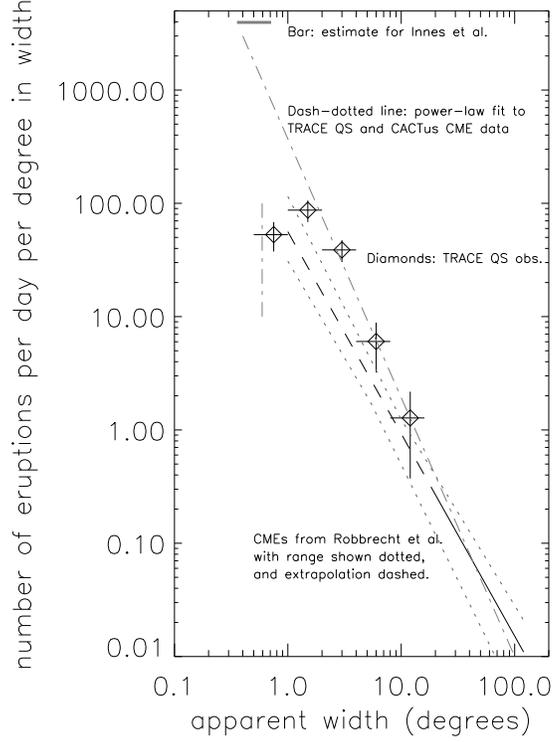}
\caption{\referee{Histogram of number of observed eruptions. The
number of events observed in the TRACE observations is shown by
diamonds, scaled to show the number of events per day per  
1-degree bin width for the apparent angular size in heliocentric degrees,
assuming a uniform
distribution across the Sun. 
The TRACE events are averaged over intervals of a factor
of two in apparent width (indicated by the horizontal bars on the
diamonds)}.  The vertical bars indicate the uncertainty in the number
based on Gaussian statistics. The solid line is the average of the
best fits to the distribution of observed coronal mass ejections
(CMEs) \referee{from Robbrecht et al.\ (2009) for 2000 to 2006; the
average power-law index and standard deviation
are $-1.78 \pm 0.17$; this best fit is extrapolated} to smaller
scales by the dashed line; the dotted lines show the envelope of
maximum and minimum values from the set of fits from Robbrecht et al.\
(2009) for the same period. To the left of the vertical dashed-dotted
line segment, the equivalent width is less than ten \referee{arcseconds,
or ten} resolution
elements in the TRACE EUV images. The grey bar near the top of the
diagram is based on the study by Innes et al.\ (2009); its scaling to
this diagram is discussed in \S~\ref{sec:discussion}. The
dashed-dotted line fits all three data sets with a power law index of
$-2.3$.}
\end{figure}\nocite{robbrecht+etal2009}\nocite{innes+etal2009}
\clearpage

\begin{figure}
\epsscale{.85}
\plotone{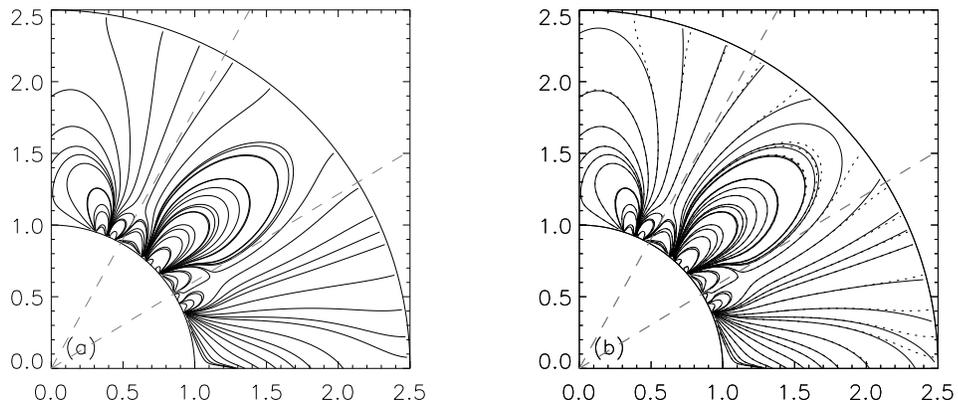}
\caption{\referee{Examples of potential field extrapolations. (a) PFSS-like
field model for test charges on the equivalent of the solar surface
(circle segment with unit radius) with an upper boundary (outer circle) 
at radial distance
of 2.5 units at which the field is forced to become radial. This
model field is invariant to rotations of 90$^\circ$. The dashed
lines enclose an extended area with the magnetic connections from the dipole
centered at 45$^\circ$ and into the open-fied domain (corresponding to the
heliosphere). (b) Same as (a) but for a potential field model in which
only the lower boundary is used. The dotted lines repeat the field lines
for the PFSS-like  model from panel (a). 
}}
\end{figure}
\clearpage

\end{document}